# High-Q Slow-Wave Coplanar Waveguides

Heng-Chia Hsu, *Student Member, IEEE*, Kaushik Dasgupta, Nathan M. Neihart, *Member, IEEE*,
Sudip Shekhar, *Member, IEEE*, Jeffrey S. Walling, *Member, IEEE*, and David J. Allstot, *Fellow, IEEE*

*Abstract*—A comprehensive study of methods of maximizing *Q* for slow-wave coplanar waveguides is described. In addition to the widths of the signal conductor and coplanar ground lines and the distance between them, the length, spacing and stacking of the metal layers of the substrate shield strips are also shown to be critical in maximizing performance. Measured results from more than 50 different devices show that a 7X increase in the quality factor (e.g., *Q* > 70 at 24 GHz in 0.18 μm CMOS) is achievable using the optimum topology with optimum dimensions.

## I. Introduction

RADIO frequency integrated circuits comprise low-loss passives and active CMOS devices with transition frequencies > 100 GHz [1]. At mm-wave frequencies (on-chip wavelengths < 10 mm or operating frequencies > 12 GHz [2]-[3]), a slow-wave coplanar waveguide is preferred to a conventional co-planar waveguide (CPW) because of its smaller size and lower losses (i.e., higher quality factor, *Q*) [2]-[7]. It also has a lower phase velocity, $v_p$, with a correspondingly shorter physical length for typical electrical lengths, *l*, of $\lambda/4$ and $\lambda/2$.

Electromagnetic interactions between the slow-wave CPW and the underlying conductive substrate should be minimized to maximize *Q*. The constituent signal and coplanar ground lines should thus be formed in the top-most (e.g., *M6*) thickest (e.g., 2.25 μm) metal layer to minimize series resistance and substrate coupling losses. A CPW is effectively edge-coupled at low frequencies; this reduces vertical coupling to the substrate and lateral coupling to adjacent passives. The skin effect dominates series resistance losses at high frequencies (e.g., 24 GHz) because the skin depth, $\delta$, for aluminum is only 0.55 μm.

Design considerations are presented for high-*Q* slow-wave coplanar waveguides for use at 24 GHz in a 6-metal 0.18 μm CMOS process. Previously published results on coplanar striplines and waveguides are extended significantly as *Q* is optimized over several topologies and the design space that includes all slow-wave CPW device dimensions.

## II. On-Chip Transmission Lines

Fig. 1(a) depicts a microstrip transmission line above a floating solid metal ground plane. The signal line is on *M6* to minimize series resistance and substrate coupling losses; *h* is ~0.75 μm (~6 μm) for the ground plane on *M5* (*M1*). A small *h* is preferred because the capacitance per unit length, *C*, of the signal line of width, $W_S$, is increased, the inductance per unit length, *L*, is decreased, and the *LC* product remains relatively constant [8]. Realization of a high characteristic impedance (e.g., $Z_o \geq 50\ \Omega$) requires a narrow signal line which has a high series loss resistance, *R*, and a low *Q* [2]. Parasitic coupling from a microstrip transmission line to nearby components also can be problematic because the electromagnetic field lines are not constrained locally.

A grounded CPW (Fig. 1(b)) [9] provides two more design degrees of freedom (*S* and $W_G$) than a microstrip line that can be used to better optimize *Q* for a given $Z_o$. *S* can be increased for higher *L* and $Z_o$, or $W_G$ can be increased for higher *C* and lower $Z_o$. Thus, a wide range of $Z_o$ values is achievable.

Even-mode excitations are usually used with the grounded (Fig. 1(b)) and slow-wave CPW (Figs. 1(c)-(d)) structures because of their symmetry with respect to the centerline of the signal conductor. As a consequence, they are less dispersive in the fundamental propagation mode than non-symmetric structures (e.g., coplanar striplines) driven by odd-mode excitations [10]. Lateral parasitic coupling is also reduced compared to a microstrip or coplanar stripline because the coplanar ground lines shield neighboring devices. The die area is therefore 30-50% less for a given degree of coupling [10].

In a grounded CPW, vertical magnetic coupling to the underlying metal ground plane induces eddy currents and opposing magnetic flux that reduces *L*. As a result, the length of the CPW needs to be increased for a given inductance. In contrast, the required length is reduced in a slow-wave CPW that uses substrate shield strips (Figs. 1(c)-(d)) to minimize the eddy current induced by currents flowing in the signal and coplanar ground lines. In a slow-wave CPW, therefore, *C* is increased, *L* remains relatively constant, and *LC* is increased, which reduces both $v_p$ and *l*.

## III. Optimization of Slow-wave CPW Structures

The area is minimized by minimizing the wavelength, $\lambda$, and the phase velocity, $v_p$,

$$\lambda = v_p / f \quad (1)$$

$$v_p = 1 / \sqrt{LC} \quad (2)$$

The characteristic impedance is

$$Z_o = \sqrt{(R + j\omega L)/(G + j\omega C)} \approx \sqrt{L/C} \quad (3)$$

where *R* (*G*) is the series (shunt) resistance (conductance) per unit length. The complex propagation constant is

Manuscript received March 4, 2010. Research supported by Semiconductor Research Corporation contract 2001-HJ-1427 and a grant from CDADIC.
H.-C. Hsu, J.S. Walling, and D.J. Allstot are with the Dept. of Electrical Engineering, Univ. of Washington, Seattle, WA 98195 (allstot@uw.edu)
K. Dasgupta is with the Dept. of Electrical Engineering, California Institute of Technology, Pasadena, CA 91125
N.M. Neihart is with the Dept. of Electrical and Computer Engineering, Iowa State University, Ames, IA 50011
S. Shekhar is with Intel Corp., Hillsboro, OR 97124

$$\gamma = \alpha + j\beta \quad (4)$$

where $\alpha$ and $\beta$ are the attenuation and phase constants per unit length, respectively. $Q$ for a transmission line resonator is [11]

$$Q = \omega \frac{\text{energy stored per unit length}}{\text{power dissipated per unit length}} = \omega \frac{E_{stored}}{P_{diss}} = \frac{\beta}{2\alpha} \quad (5)$$

where $E_{stored} = W_m + W_e$ and $W_m$ ($W_e$) is the stored magnetic (electric) energy. $P_{diss}$ includes: (1) $P_{ohmic}$—power dissipated due to the metal series resistance, skin and proximity effects and electromagnetic coupling (e.g., eddy current) losses; (2) $P_{dielectric}$—power consumed because of dielectric losses.

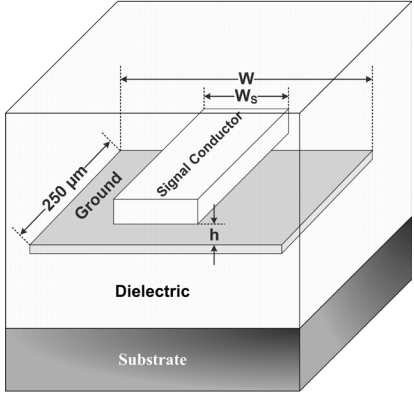

(a)

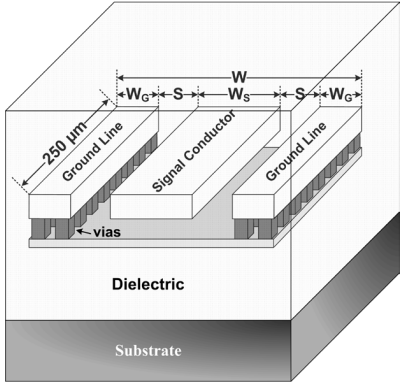

(b)

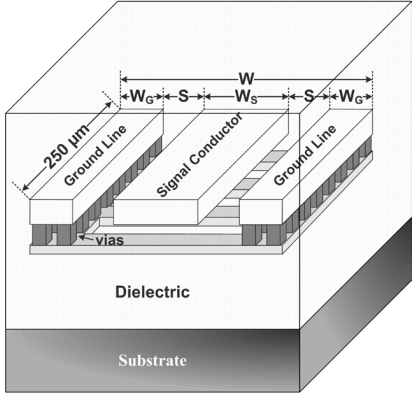

(c)

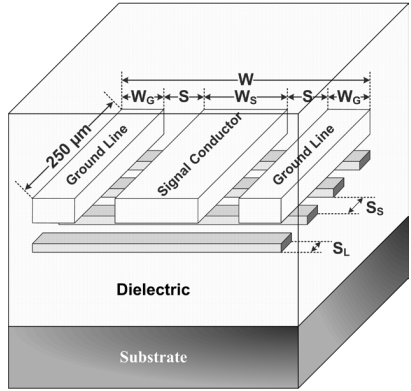

(d)

Fig. 1. (a) Microstrip line; (b) grounded coplanar waveguide; slow-wave coplanar waveguides with (c) grounded and (d) floating substrate shield strips.

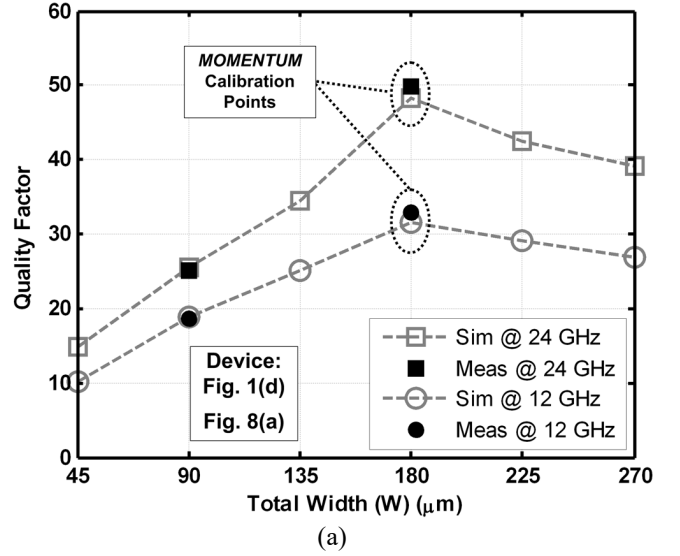

(a)

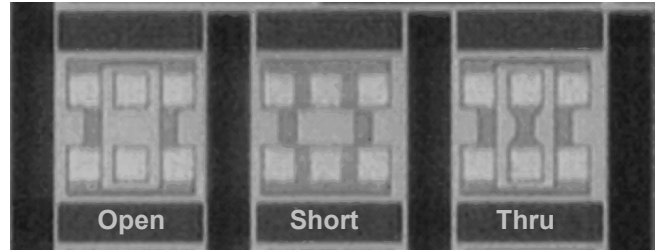

(b)

Fig. 2. (a) Measured and simulated $Q$ values @ 12 and 24 GHz vs. $W$ with $S_L = S_S = 3$ μm. The other dimensions (Table I) are optimized for maximum $Q$ with $Z_o \sim 40$ Ω. (b) First, SOLT calibration to the probe heads was performed using an off-chip calibration standard [18]. Next, on-chip de-embedding was performed wherein the pad capacitance and inductance were measured using the on-chip *short* and *open* structures above. Finally, the inductance of the launch into the DUT was de-embedded using the on-chip *thru* structure. Note: All transmission lines herein are 250 μm long.

TABLE I
SLOW-WAVE CPW DIMENSIONS ($Z_O \sim 40$ Ω)
DEVICE: FIGS. 1(d) AND 8(a)

| Total W (μm) | $W_S$ / S / $W_G$ (μm) |
|---|---|
| 45 | 10 / 7.5 / 10 |
| 90 | 28 / 21 / 10 |
| 135 | 18 / 38.5 / 20 |
| 180 | 20 / 55 / 25 |

| 225 | 22 / 76.5 / 25 |
| 270 | 24 / 98 / 25 |

### A. Slow-wave CPW Performance vs. W

Measured and simulated values of $Q$ vs. $W$ for the slow-wave CPW of Figs. 1(d) and 8(a) are given in Fig. 2[1]. The dimensions of the floating substrate shield strips are fixed at $S_L = S_S = 3$ μm whereas $W_S$, $W_G$ and $S$ are optimized to achieve a maximum $Q$ for each width with $Z_o \sim 40$ Ω (Table I). The overall maximum $Q$ is obtained with $W = 180$ μm in the 0.18 μm CMOS process. The effects on $Q$ for each of the dimensions ($W_S$, $W_G$, $S$, $S_L$ and $S_S$) are examined next with $W = 180$ μm in all cases.

### B. Slow-wave CPW Performance vs. $W_S$, $W_G$, and $S$

Variations in the dimensions $W_S$, $W_G$ and $S$ impose tradeoffs among $Q$, $v_p$ and $Z_o$. As $W_S$ and $W_G$ are increased from small sizes, $Q$ first increases because the series resistance and skin-effect losses decrease but it then decreases for larger widths because the eddy current and dielectric losses increase.

Induced eddy current is found using Faraday's Law

$$\oint_C E \cdot dl = -\frac{d}{dt} \iint_S B \cdot ds \quad (6)$$

wherein a changing magnetic flux, $\Phi$, through the surface of a metal line, $s$, induces an electromotive force, $V$, in the metal line with closed contour, $c$. Hence,

$$V = -\frac{d\Phi}{dt} \quad (7)$$

The eddy currents induced in the substrate and shield strips create an opposing magnetic flux. The effect is greater at higher frequencies and in wider metal lines because of the increased surface area. The attenuation constant is [12]

$$\alpha \approx 0.5(R\sqrt{C/L} + G\sqrt{L/C}) = 0.5(R/Z_0 + GZ_0) \quad (8)$$

An increase in $W_S$ reduces the series resistance and skin effect losses (i.e., $R$) but increases the eddy current (i.e., $G$ and substrate loss) and dielectric losses; the latter occurs because more field lines pass through the dielectric layer. An optimum $Q$ versus $W_S$ therefore exists (Fig. 3(a)). (Similar phenomena for coplanar stripline structures were reported by Andress, et al. [13] and Krishnaswamy, et al. [14].) Because $C$ increases faster than $L$ decreases, $v_p$ (Fig. 3(b)) and $Z_o$ (Fig. 3(c)) decrease as $W_S$ increases. Similar results were described by Sayag, et al. [7].

An increase in $W_G$ also reduces the series resistance and skin effect losses but increases the dielectric and eddy current losses. Thus, an optimum $Q$ versus $W_G$ also exists (Fig. 4(a)). As $W_G$ is increased, the fringing capacitance increases but the effects on $v_p$ (Fig. 4(b)) and $Z_o$ (Fig. 4(c)) are less than for increased $W_S$ because fringing capacitance increases sub-linearly versus $W_G$.

---

[1] The *ADS MOMENTUM* simulator is calibrated to account for process variations, model inaccuracies, etc. The metal and substrate and definitions are provided by the vendor, but typically the substrate conductivity is not. Starting at an initial guess of 10 S/m, it is adjusted empirically to achieve a good match between the simulated and measured s-parameters at both 12 GHZ and 24 GHz as shown in Fig. 2 for the slow-wave CPW with W = 180 μm. All subsequent simulations are performed after these one-time calibrations.

Flux lines that link the signal and coplanar ground lines cause lateral coupling that can be modeled as a mutual inductance

$$M = \Phi_{couple} / I \quad (9)$$

An increase in $S$ reduces $M$ and the losses due to inductive coupling and the proximity effect. Hence, $L$, $W_m$, $Q$ and $Z_o$ increase and $v_p$ decreases. With further increases in $S$ a greater percentage of the return current flows in the substrate. Therefore, substrate- and shield-related losses increase and $Q$ decreases. The measured results of Fig. 5 confirm these effects.

$Q$ is increased by 5X with optimum $W_S$, $W_G$ and $S$ dimensions at 24 GHz as demonstrated in Figs. 3-5. These optimum sizes confirm the basic design trade-off between $W$ and $Q$ depicted in Fig. 2. Analogous results for $v_p$ versus $W_S$ and $S$ in a coplanar stripline at 60 GHz were reported by LaRocca, et al. [15].

The induced voltage, $V = -j\omega MI$, is found from (7). Current flowing ($I_S$) in the signal line generates voltages on the coplanar ground lines, and the resulting currents ($I_G$) induce a voltage, $V_S$, on the signal line. These interactions can be modeled as a reflected resistance ($Z_r$) [16]

$$Z_r = \frac{V_S'}{I_S} = \frac{-j\omega M}{-(I_S - I_{eddy})} \cdot I_G = \frac{-j\omega M}{-(I_S - I_{eddy})} \cdot \left(\frac{-j\omega M \cdot I_S}{R_G + j\omega L_G}\right)$$

$$= \frac{(\omega M)^2}{R_G + j\omega L_G} \cdot \frac{I_S}{I_S - I_{eddy}} \quad (10)$$

The $(I_S - I_{eddy})$ term accounts for the eddy current flow opposite to $I_S$ near the middle of signal conductor (Fig. 6(a)). The impedance of the signal (ground) conductor is modeled as a resistance, $R_S$ ($R_G$) in series with an inductance, $L_S$ ($L_G$) (Fig. 6(b)) plus $Z_r$

$$Z_S = (R_S + j\omega L_S) + \frac{(\omega M)^2}{R_G + j\omega L_G} \cdot \frac{I_S}{I_S - I_{eddy}}$$

$$= R_S + R_G \left[\frac{(\omega M)^2}{R_G^2 + (\omega L_G)^2} \cdot \frac{I_S}{I_S - I_{eddy}}\right]$$

$$+ j\omega L_S - j\omega L_G \left[\frac{(\omega M)^2}{R_G^2 + (\omega L_G)^2} \cdot \frac{I_S}{I_S - I_{eddy}}\right] \quad (11)$$

Owing to the lateral mutual inductance, $M$, and $I_{eddy}$, the effective resistance (inductance) of the signal conductor is increased (decreased) and both effects degrade $Q$. It is clear from (11) that eddy current losses appear as resistive losses. Because the proximity effect adds to $P_{ohmic}$, increases in $W_S$ and $W_G$ have a diminished effect in reducing ohmic losses. This is because of the increased current density near the inner edges. This drawback is addressed by using a slow-wave CPW with stacked metal layers.

### C. Slow-wave CPW Performance vs. $S_L$ and $S_S$

Grounded (Fig. 1(c)) or floating (Fig. 1(d)) substrate shield strips enhance the slow-wave effect. They also reduce eddy currents in the substrate; i.e., some flux lines link through the spaces between the strips to the substrate, whereas others ($B_T$) also link, but are attenuated by the strips (Fig. 6(c)). The former linkages increase as $S_S$ increases whereas the latter decrease with more substrate shield layers. Eddy currents induced in the shield strips create an opposing magnetic flux, $B_{eddy}$. This effect increases as $S_L$ increases. $B_{eddy} = B_S - B_T$ should thus be minimized to maximize the stored magnetic energy. Hence, there exist optimum $S_L$ and $S_S$ values that maximize $Q$.

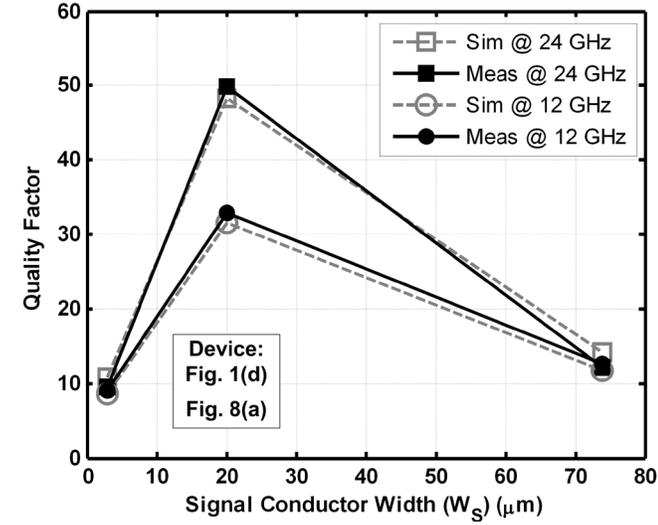

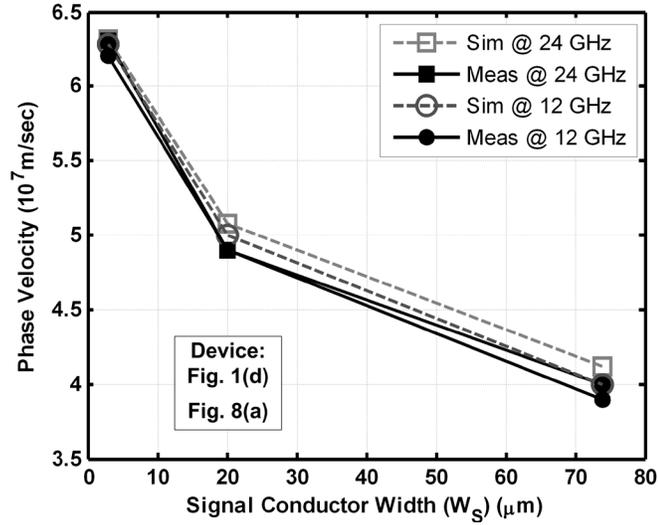

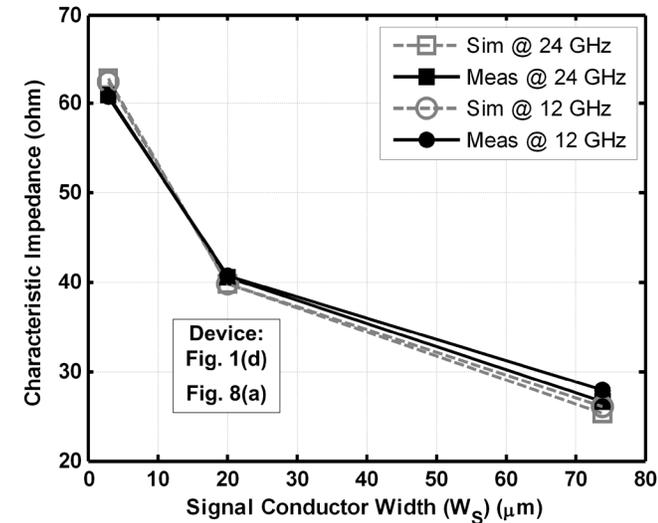

Fig. 3. (a) $Q$, (b) $v_p$ and (c) $Z_o$ @ 12 GHz and 24 GHz versus $W_S$ for device 4F in Table III.

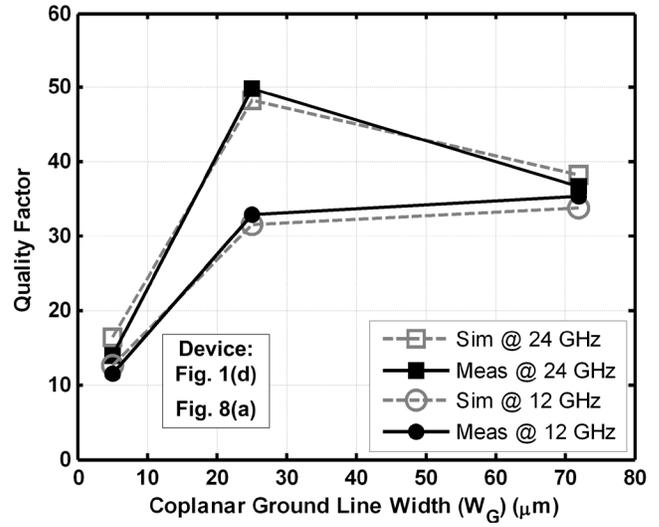

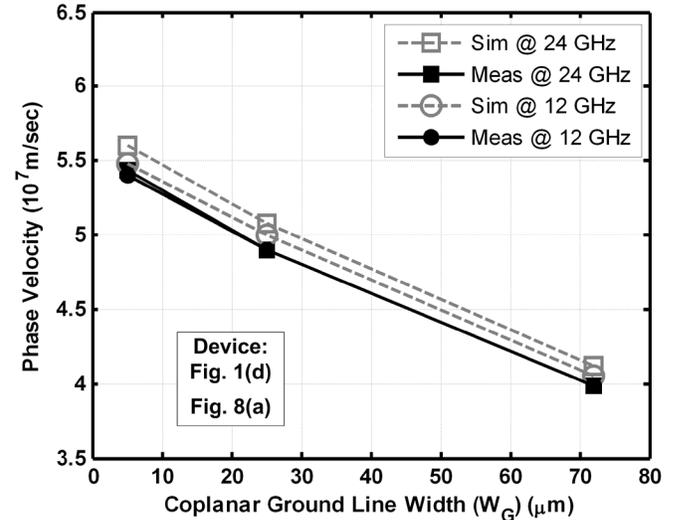

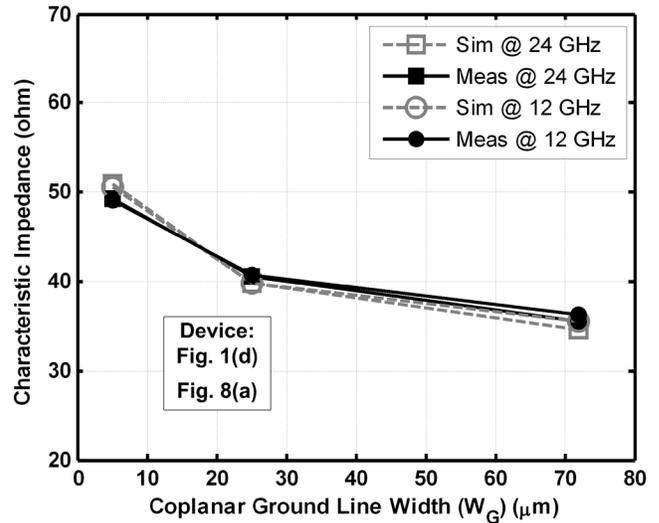

Fig. 4. (a) $Q$, (b) $v_p$ and (c) $Z_o$ @ 12 GHz and 24 GHz versus $W_G$ for device 4F in Table III

Consider (11) with $R_G$ and $L_G$ replaced by $R_{ST}$ and $L_{ST}$, the resistance and inductance of a substrate shield strip (Fig. 6(b)). Inductive coupling (i.e., the last term in (11)) is decreased by increasing $R_{ST}$ and $L_{ST}$. The vertical magnetic coupling between the signal and coplanar ground lines and the substrate shield strips, $M_T$, replaces $M$ and is constant because the vertical spacing is fixed. Both $L_{ST}$ and $R_{ST}$ increase as $W$ increases and $S_L$ and $S_T$ decrease (Fig. 6(c)) [17]

$$L_{ST} = 2\mu_o W \left[ \ln\left(\frac{2W}{S_L + S_T}\right) + 0.5 \right]$$

(12)

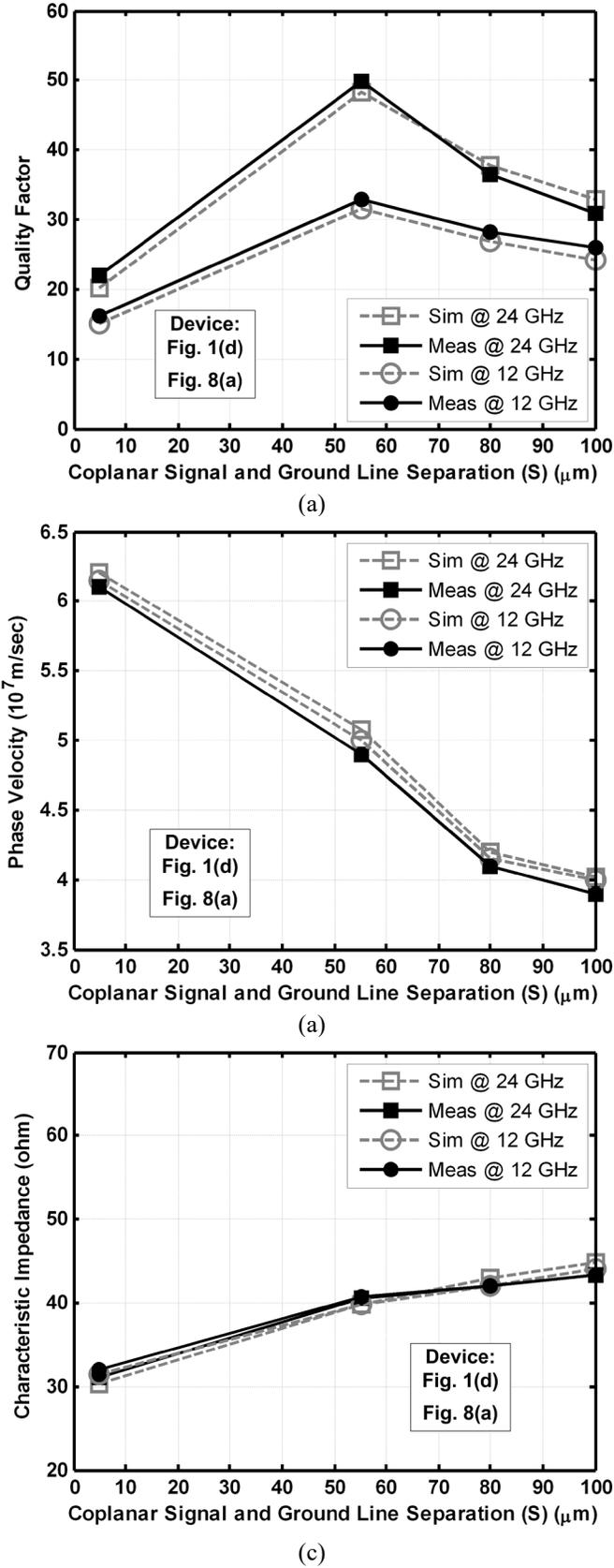

Fig. 5. (a) $Q$, (b) $v_p$ and (c) $Z_o$ @ 12 GHz and 24 GHz versus $S$ for device 4F in Table III.

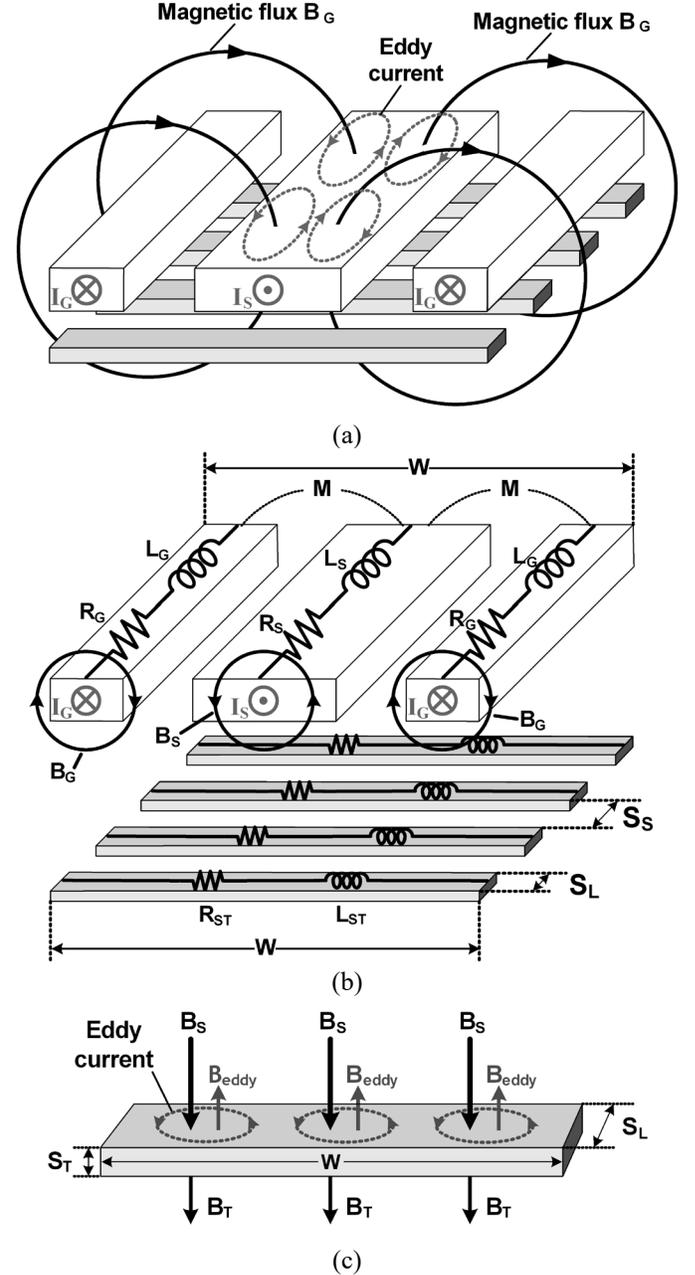

Fig. 6. (a) Lateral inductive coupling from the coplanar ground lines causes current crowding at the edges of the signal line; (b) lumped-element model for magnetic coupling, and (c) induced eddy currents and magnetic flux attenuation through a substrate shield strip.

From (11), therefore, the total resistance (inductance) decreases (increases). These characteristics suggest the use of a large $W$ with minimum $S_L$ and $S_T$ values. As the conductivity of the substrate shield strips increases, however, the induced eddy currents and associated losses increase. For a given length (e.g., 250 μm) with minimum $S_L$ ($S_S$), eddy currents induced in the substrate (substrate shield strips) increase as $S_S$ ($S_L$) increases because of the increased flux linkages. Thus, there exists an optimum $S_L/S_S$ combination for maximum $Q$ (Fig. 7). The 1/1 and 1/3 designs give maximum $Q$ values of 60 and 75, respectively. Eddy current is proportional to conductivity; that of aluminum (~$3.5 \times 10^7$ S/m) is much greater than that of the substrate (~10 S/m). As the area of the shield strip decreases, a smaller fraction of the flux lines impinge on it, so less opposing magnetic flux $B_{eddy}$ is generated and $Q$ is increased. In contrast, the maximum $Q$ is about only 30 for the 1/11 device because its eddy current losses in the substrate are greater. The substrate shield of the 35/1 structure, which approximates the ground plane of the grounded CPW in Fig. 1(b), gives the lowest maximum $Q$ value; this occurs because it is most vulnerable to eddy current losses in the highly conductive shield strips. A slow-wave CPW with minimum $S_L$ is therefore preferred for maximum $Q$.

TABLE II
SLOW-WAVE CPW DIMENSIONS ($Z_o \sim 40\ \Omega$)
DEVICE: FIGS. 1(d) AND 8(a)

| $W_S / S / W_G$ (μm) | $S_L / S_S$ (μm) |
|---|---|
| 20 / 55 / 25 | 1 / 1 |
| 21 / 54.5 / 25 | 1 / 3 |
| 35 / 47.5 / 25 | 1 / 11 |
| 19 / 55.5 / 25 | 3 / 1 |
| 20 / 55 / 25 | 3 / 3 |
| 30 / 50 / 25 | 3 / 11 |
| 12 / 59 / 25 | 35 / 1 |
| 13 / 58.5 / 25 | 35 / 3 |
| 15 / 57.5 / 25 | 35 / 11 |

The phase velocities of the $S_L = 1$ μm and 3 μm devices show a similar trend (Fig. 7(b)). As $S_S$ decreases with $S_L$ fixed, $C$ increases, but the effective inductance is approximately $L_S$ in (11) assuming $L_{ST}$ and $R_{ST}$ are large and $B_{eddy}$ is small; hence, $v_p$ decreases according to (2). For the $S_L = 35$ μm devices, however, as $S_S$ increases, the effective inductance increases faster than $C$ decreases so $v_p$ decreases.

Different $S_L$ and $S_S$ dimensions give different $Z_o$ values for fixed $W_S$, $S$ and $W_G$ [18]. For $Z_o \sim 40\ \Omega$, the 1/3 design yields the overall maximum $Q$ (Fig. 7(a)).

### D. Slow-wave CPW Performance vs. Metal Stack Thicknesses

Metal thickness studies for on-chip spiral inductors have shown that metal layer stacks reduce losses and increase $Q$ [19]-[21]. Moreover, the effects of stacked metals on the $R$, $C$, $L$ and $Z_o$ characteristics of conventional CPW structures are described by Woods, et al. in [22].

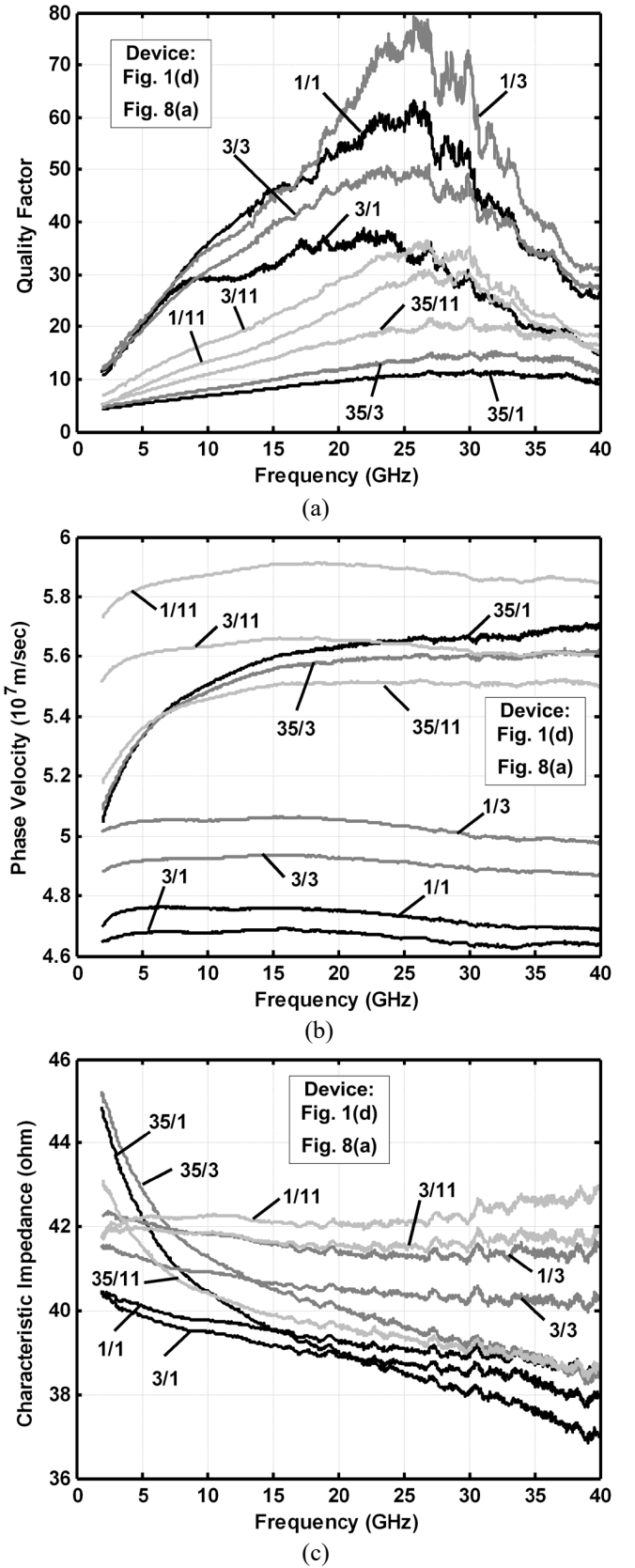

Fig. 7. (a) $Q$, (b) $v_p$, and (c) $Z_o$ for different $S_L/S_S$ values. Device dimensions are given in Table II.

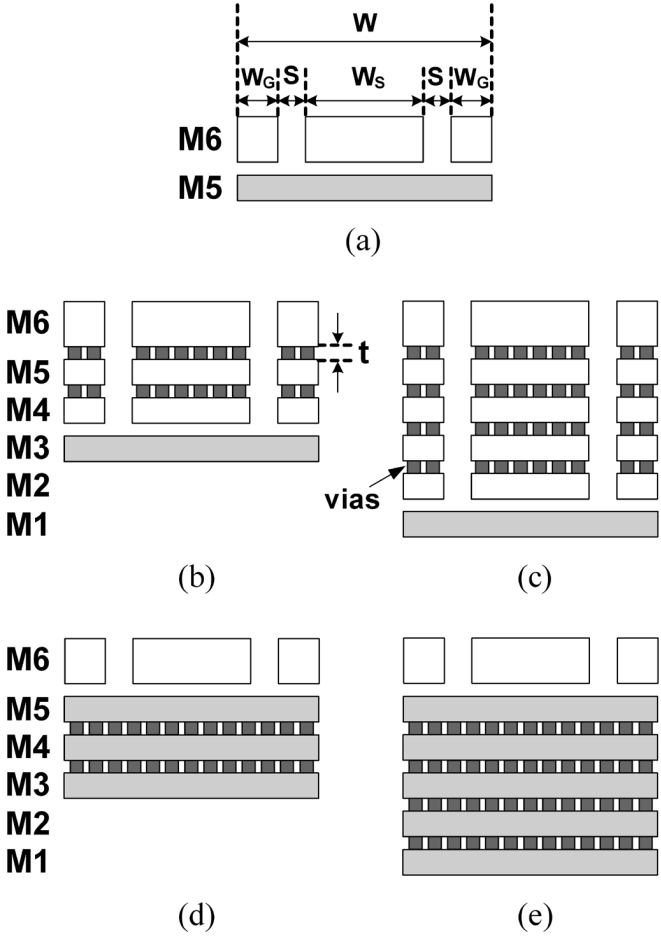
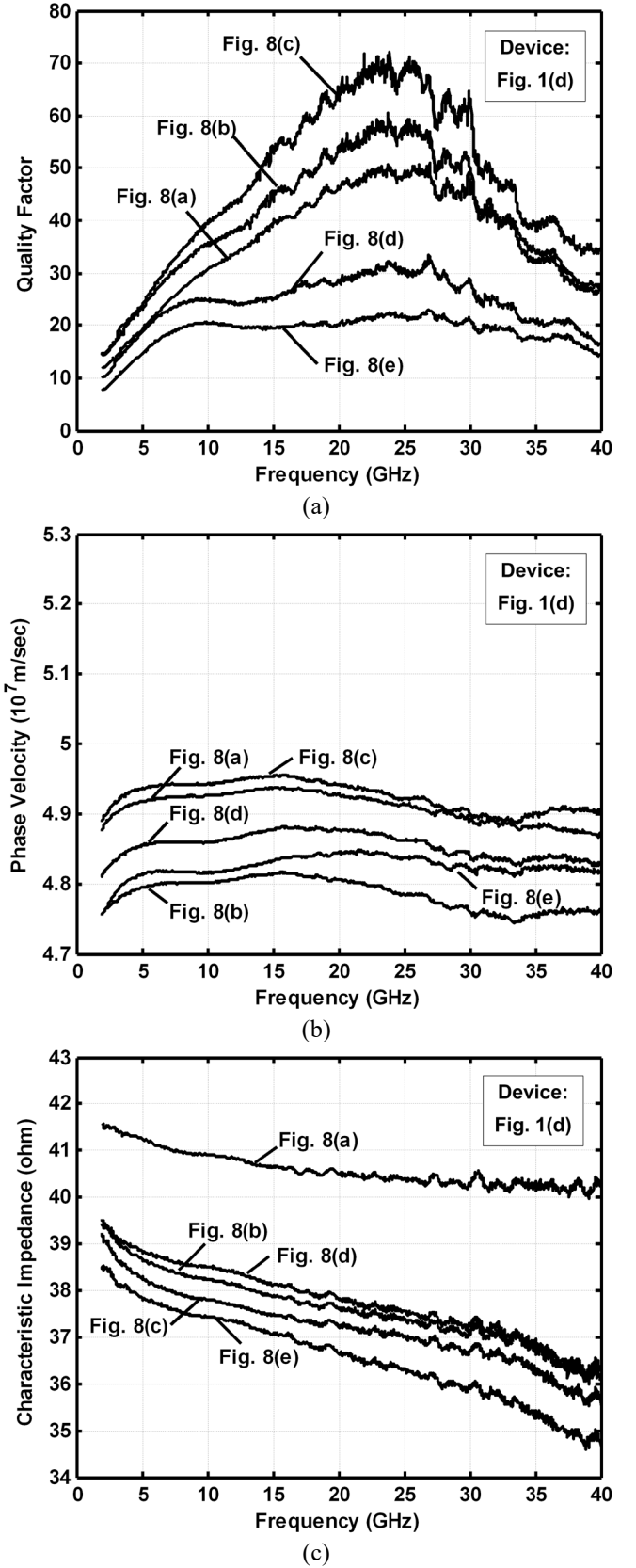

Fig. 8. Slow-wave CPW structures with different signal, ground, and substrate shield metal stacks in a 6-layer metal 0.18 μm CMOS process. (white = signal and ground metals; gray = substrate shield metals and dark gray = vias)

The effects of metal stacks on the key performance parameter, $Q$, as well as $v_p$ and $Z_o$ of the slow-wave CPW structures are now examined. In the 0.18 μm CMOS process, the signal and coplanar ground lines are formed in the thick (~ 2.25 μm) top metal layer (Figs. 8(a)(d)(e)) only, or with additional metal layers (thickness ~ 0.5 μm) connected in parallel using vias (thickness ~ 0.75 μm) (Figs. 8(b)(c)). Single-layer substrate shield strips should occupy the highest metal layer immediately below the signal and coplanar ground stack (Figs. 8(a)-(c)), or include multiple lower metal layers (Figs. 8(d)(e)). A stack approximates a single metal layer of thickness equal to the sum of the constituent thicknesses. For fixed $W_S$, $W_G$ and $S$, the signal metal stack has lower $R$, smaller $L$, and higher sidewall capacitance. The capacitance between the signal stack and the underlying substrate shield or shield stack is approximately constant for all five structures in Fig. 8. Hence, the relative passive element values of the top metal stacks (Figs. 8(a)-(c)) are

$$C_a < C_b < C_c$$
$$L_a > L_b > L_c$$
$$R_a > R_b > R_c \tag{13}$$

Fig. 9. (a) $Q$, (b) $v_p$, and (c) $Z_o$ values for the different metal stack configurations with floating substrate shield strips. All have the same design dimensions: $W = 180$ μm, $W_S = 20$ μm, $W_G = 25$ μm, $S = 55$ μm, $S_L = 3$ μm and $S_S = 3$ μm.

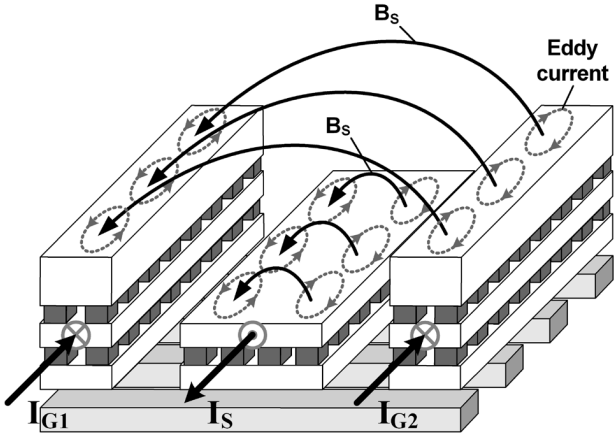

Fig. 10. The metal stack of Fig. 8(b) with floating substrate shield strips and with *M6* and its associated vias cut away from the signal line. Magnetic flux lines due to $I_S$ and the corresponding induced eddy currents are depicted.

$L$ and $C$ vary in opposite directions, so $Z_o$ decreases as the thicknesses of the signal and coplanar ground stacks increase (Fig. 9(c)). This also suggests, from (2), a minimum $v_p$ as confirmed in Fig. 9(b). It is achieved for the topology of Fig. 8(b); the smallest slow-wave effect is obtained for the device of Fig. 8(c). The capacitance between the signal stack and substrate shield dominates over the side-wall capacitance and the inductance is relatively constant [22]. Therefore, $v_p$ and $Z_o$ are approximately constant in Figs. 8(a)-(c).

Magnetic flux ($B_S$) is created by the current, $I_S$, that flows through the signal stack in Fig. 10 decreases exponentially from $H_{in}$ at the entry point on the surface versus penetration distance, $d$ [23]

$$H_t = H_{in} e^{-d/\delta_S}$$
(14)

where $\delta_S$ is the skin depth,

$$\delta_S = \sqrt{2/\omega\mu\sigma} \quad (15)$$

and $\mu$ and $\sigma$ are the permeability and conductivity, respectively. Flux lines due to the return currents, $I_{G1}$ and $I_{G2}$, are not shown.

The current density is uniform in the metal stacks at low frequencies because $\delta_S$ is large. At high frequencies, however, current crowding occurs because $\delta_S$ is small and the eddy currents flow in the opposite (same) direction as $I_S$ near the middle (edges) of the signal stack (Fig. 10). Similarly, flux lines that impinge on the vertical areas of the signal stack facing the ground stacks cause current crowding to the top or bottom surfaces. Ohmic loss resistance increases, therefore, as the skin depth decreases. This effect is mitigated using signal stacks with longer peripheries; as a result, the measured peak $Q$ values (Fig. 9(a)) corresponding to Figs. 8(a)-(c) range from 50 to 70.

The performance with different substrate shield stacks (Figs. 8(a)(d)(e)) is studied with both $S_L$ and $S_S$ equal to 3 μm. The relative inductances and resistances for Figs. 8(a)(d)(e) are

$$L_a > L_d > L_e$$
$$R_a > R_d > R_e$$
(16)

Although resistance is reduced using a metal stack, ohmic losses due to the skin effect are not significantly impacted. This is due to the fact that currents flow mostly in the signal and coplanar ground stacks. The magnetic field that penetrates through the thick shield stack to the substrate is attenuated, which corresponds to increased absorption from (14) and (15).

From (5), $Q$ is proportional to ($W_m + W_e$) and inversely proportional to $P_{diss}$. As the thickness of the substrate shield stack increases, the absorption increases and the inductance and corresponding $W_m$ decrease. In contrast, $W_e$ increases because of the increased fringing capacitance between the substrate shield stack and signal and coplanar ground lines. Therefore, dielectric loss is also increased. Consider (11) with $R_G$ and $L_G$ replaced by $R_{ST}$ and $L_{ST}$. Both are decreased (16) using the shield stack so that inductive coupling is increased. Eddy current losses also increase as the resistance of substrate shield stack decreases.

The expected lower $Q$ (from (5)) for the slow-wave CPW of Fig. 8(e) is confirmed in Fig. 9(a). It is also clear from (12) that thin substrate shield strips are preferred. This analysis also explains the $v_p$ and $Z_o$ characteristics in Figs. 9(b)(c). Therefore, the preferred slow-wave CPW structure is a thick signal and coplanar ground stack over a single substrate shield layer.

### E. Other Substrate Shield Considerations

The characteristics of the slow-wave CPW of Fig. 8(a) with grounded (Fig. 1(c)) and floating (Fig. 1(d)) substrate shield strips are compared in this section.

Fig. 11(a) shows a conventional CPW with the signal and ground lines excited by AC sources [3]. Losses are incurred because of coupling to the conductive substrate. The voltage at A, the surface of the substrate, is determined by the RC voltage dividers that model the dielectric ($R_{SiO2}$ // $C_{SiO2}$) and substrate ($R_{SUB}$ // $C_{SUB}$) layers. $R_{SiO2}$ and $R_{SUB}$ are much greater than the reactances $(\omega C_{SiO2})^{-1}$ and $(\omega C_{SUB})^{-1}$ at high frequencies; hence,

$$V_A \simeq \frac{C_{SiO2}}{C_{SiO2} + C_{SUB}} V$$
(17)

An ideal ground shield between the signal and coplanar ground lines and the substrate eliminates this coupling.

The use of a grounded slotted substrate shield in spiral inductors to reduce coupling to the substrate is described by Yue, et al. [24]. Slow-wave CPW structures, with both grounded and floating substrate shield strips, are also considered by Komijani, et al. [5], Cheung, et al. [2]-[3] and Shi, et al. [6]. Figs. 11(b)-(c) depict electric field lines for the two cases. Most of the field lines contribute to vertical capacitances but some contribute to lateral fringing capacitances. Due to the bi-directional coupling, the electric field lines are twice as long nominally for the device with the floating substrate shield [3]. $C_{G1}$ ($C_{G2}$) in Fig. 11(b) should therefore be larger than $C_{F1}$ ($C_{F2}$) in Fig. 11(c). From (17), with $C_{SiO2}$ replaced with $C_{F2}$ and $C_{G2}$ for the two cases, respectively, the coupling loss in the floating shield case is less than in the grounded shield device. When $V$ is positive, the charges on the floating substrate shield are distributed as shown; when $V$ cycles negative, the signs are reversed. Consequently, there is a time-varying lateral electric field in the floating shield layer itself that prevents it from acting as an ideal grounded substrate shield layer [3].

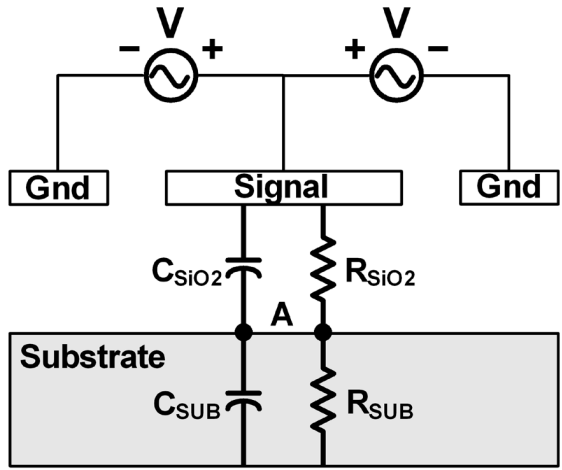

(a)

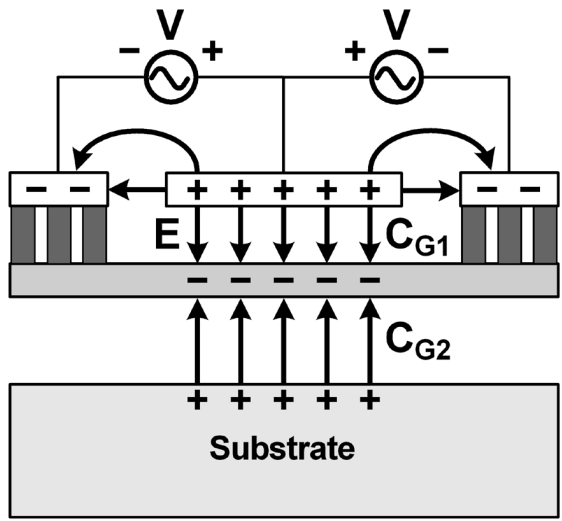

(b)

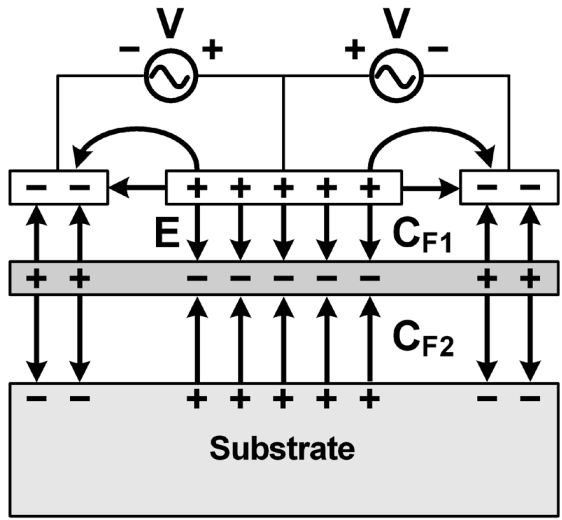

(c)

Fig. 11. (a) Conventional CPW and slow-wave CPW with (b) grounded and (c) floating substrate shield strips.

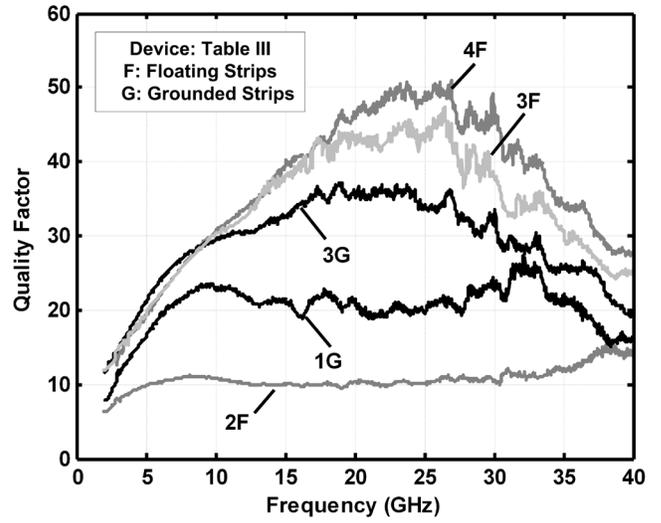

(a)

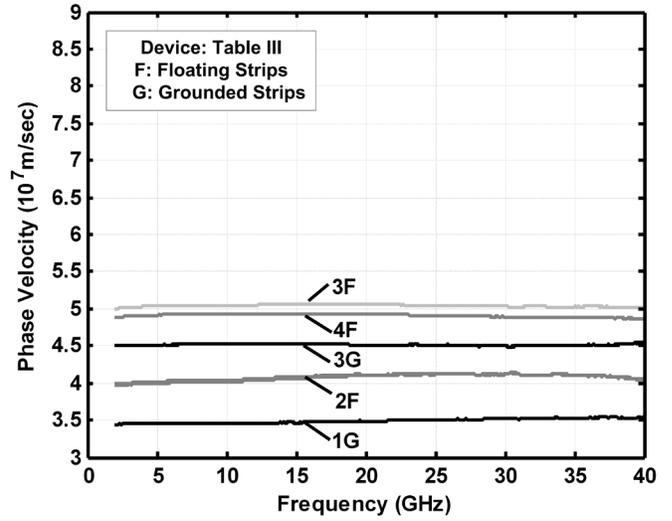

(b)

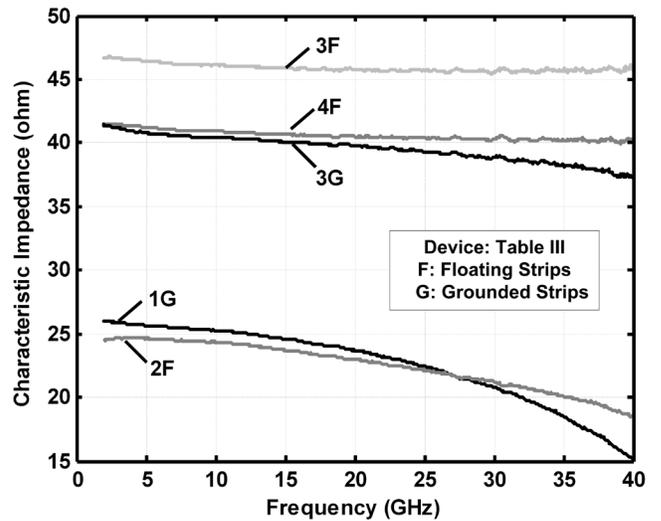

(c)

Fig. 12. The effects of grounded vs. floating substrate shield strips on (a) $Q$, (b) $v_p$, and (c) $Z_o$ with $W = 180$ μm. Device dimensions are given in Table III.

TABLE III
SLOW-WAVE CPW DEVICES
FLOATING SUBSTRATE SHIELDS: FIGS. 1(d) AND FIG. 8(a)
GROUNDED SUBSTRATE SHIELDS: FIGS. 1(c) AND FIG. 8(a)

| Slow-wave CPW | Substrate Shield | $W$ (µm) | $W_S / S / W_G$ (µm) | $S_L / S_S$ (µm) | $Z_o$ (Ω) |
|---|---|---|---|---|---|
| 1G | Grounded | 180 | 35 / 45.5 / 27 | 3 / 1 | ~25 |
| 2F | Floating | 180 | 74 / 26 / 27 | 3 / 1 | ~25 |
| 3G | Grounded | 180 | 16 / 57 / 25 | 3 / 3 | ~40 |
| 4F | Floating | 180 | 20 / 55 / 25 | 3 / 3 | ~40 |
| 5G | Grounded | 90 | 34 / 18 / 10 | 3 / 4 | ~25 |
| 6F | Floating | 90 | 35 / 7.5 / 20 | 3 / 1 | ~25 |
| 7G | Grounded | 90 | 17 / 26.5 / 10 | 3 / 3 | ~40 |
| 8F | Floating | 90 | 28 / 21 / 10 | 3 / 3 | ~40 |
| 3F | Floating | 180 | 16 / 57 / 25 | 3 / 3 | ~46 |
| CPW | No Shield | 90 | 35 / 7.5 / 20 | N/A | ~46 |

Consider devices 3G and 3F in Table III with identical dimensions and grounded and floating substrate shield strips, respectively. Because the latter has less coupling loss and capacitance, $Q$ (5), $v_p$ (2), and $Z_o$ (3) are all higher (Fig. 12). Hence, a device with a floating (grounded) substrate shield is most suitable for high (low) $Z_o$ designs. Next, consider devices 1G and 2F where both are designed for $Z_o \sim 25$ Ω. The latter requires a much wider $W_S$ for a similar $C$, which leads to increased eddy current, coupling and dielectric losses. $Q$, $v_p$, and $Z_o$ are thus all lower as shown. Finally, consider devices 3G and 4F with both designed for $Z_o \sim 40$ Ω. The wider $W_S$ in 4F reduces series resistance and skin effect losses compared to 3G that has a narrower $W_S$ so that $Q$, $v_p$, and $Z_o$ are generally higher.

The low impedance devices 5G and 6F with $W = 90$ µm give small peak $Q$ values (Fig. 13(a)). From (3), high capacitance is needed to achieve $Z_o \sim 25$ Ω. Because she signal conductor widths are similar (Table III), however, the needed approximately equal capacitances are obtained for 5G with $S_L/S_S = 3/4$ and 6F with $S_L/S_S = 3/1$. It is clear from Figs. 12(a) and 13(a) that higher peak $Q$ values are achieved for devices with larger $W$ values. The high impedance devices with floating substrate shields give the highest peak $Q$ values for a given $W$.

$C_{SiO2}$ in Fig. 11(a) is approximately equal to $C_{G2}$ in Fig. 11(b); a slow-wave CPW with grounded substrate shield strips, therefore, has capacitive coupling comparable to a conventional CPW. The latter experiences substantial eddy current losses in the conductive substrate due to the absence of shield strips. Fig. 13 includes measured results for a conventional CPW (Table III) with $Z_o \sim 46$ Ω. Compared to device 7G with $Z_o \sim 40$ Ω, it gives lower $Q$ and higher $v_p$. An even lower value of $Q$ is expected if a wider $W_S$ is used to obtain $Z_o \sim 40$ Ω. Additionally, by comparing 1/11 in Fig. 7 with the CPW in Fig. 13 with the same $W_S$ and similar $W_G$ values (Tables II and III), it is seen that the former has a lower $v_p$ due to the capacitance of the substrate shield strips. That is, with larger $C$, larger $S$ and larger $L$, and assuming $R_{ST}$ and $L_{ST}$ are large the effective inductance is approximately equal to $L_S$ in (11).

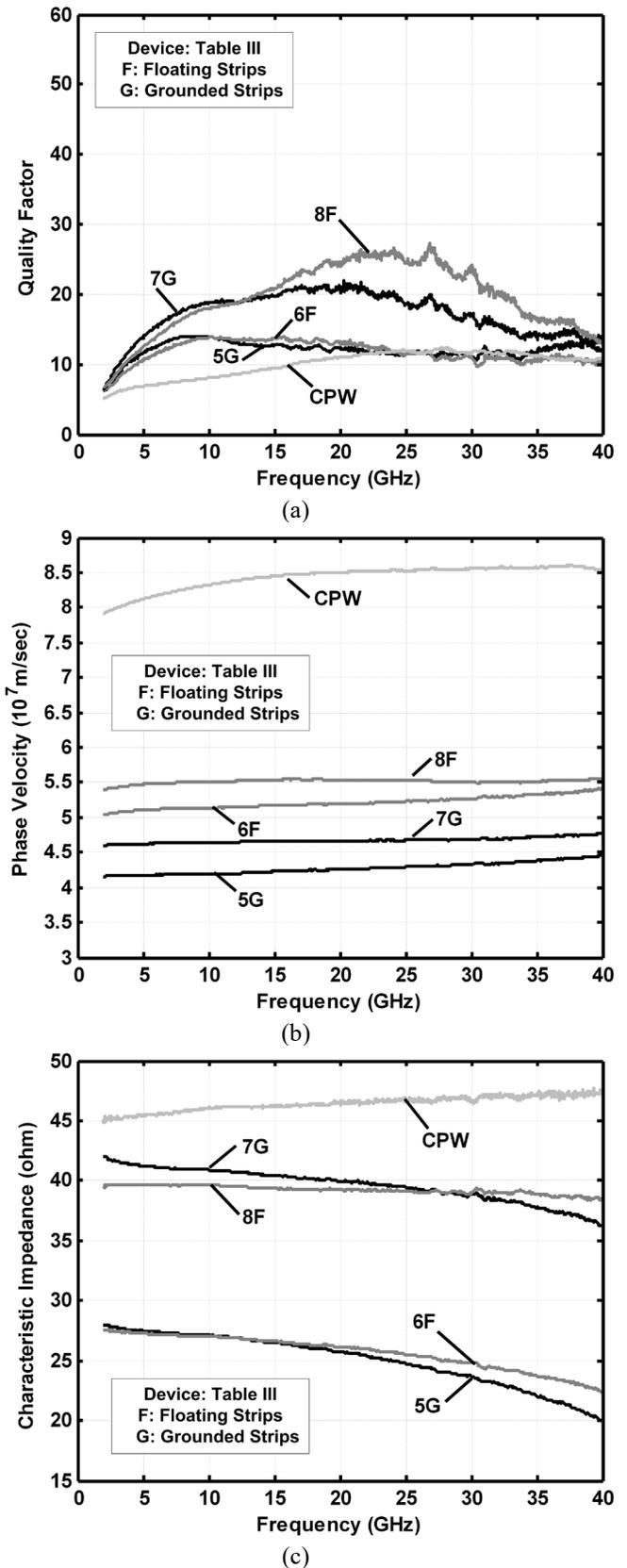

Fig. 13. The effects of grounded vs. floating substrate shield strips on (a) $Q$, (b) $v_p$, and (c) $Z_o$ with W = 90 µm. Device dimensions are given in Table III.

Slow-wave CPW structures with overlapping and offset substrate shield stacks are shown in Figs. 14(a) and (c), respectively. In Fig. 14(a), layers *M1-M5* are stacked in an overlapped pattern that approximates a single thick metal layer and attenuates the magnetic field that impinges on the conductive substrate strips (Fig. 14(b)). Some magnetic flux lines couple directly between the signal stack and the substrate. In Fig. 14(c), layers *M1-M5* are stacked in an alternating offset pattern in order to fill in the gaps ($S_S$) of the metal substrate strip layers immediately above and below so that magnetic flux lines do not link to the substrate.

Fig. 14(b) depicts the flux densities associated with the device with overlapped shield strips. $B_S$ first penetrates *M5* where it induces eddy current and opposing flux ($B_{eddy}$). The net magnetic flux, ($B_S - B_{eddy}$) then impinges on *M4*, and so on. This effect accounts for an increase in eddy current losses, a decrease in the stored magnetic energy and a decrease in *Q*. Fig. 14(d) shows the flux associated with the device with offset shield strips. The magnetic field intensity is attenuated similarly by the multiple substrate shield layers. However, greater opposing magnetic flux is induced in the substrate strips because more metal area is exposed to $B_S$ in the offset arrangement.

The overlapping and offset arrangements can be used with grounded or floating substrate shield stacks or a combination of the two. Fig. 15(a) plots *Q* for the first five devices in Table IV with grounded substrate shields. Conclusions drawn from the measurements are: (1) The single grounded substrate shield provides the highest peak *Q*; (2) the overlapping arrangement yields a higher peak *Q* than the devices with the offset substrate shields for the same number of layers; (3) the peak *Q* is lower when more overlapping or offset shield layers are used.

Fig. 15(b) plots *Q* for the next five entries in Table IV with floating substrate shields. The peak *Q* values are higher but the overall trends are the same as in Fig. 15(a). For the last two hybrid substrate shield entries in Table IV, Fig. 15(c) shows that the peak *Q* is higher for the overlapping shield configuration, and generally that there is no advantage in using a device with a hybrid substrate shield stack.

Fig. 16 shows measured $v_p$ values for the slow-wave CPW devices detailed in Table IV. With multiple substrate shield layers, the fringing and other stray capacitances are increased, but the inductance is decreased due to the attenuation of the magnetic fields. Hence, $v_p$ with multiple substrate shields of the same type is similar to that of the corresponding device with a single substrate shield layer. The results of Fig. 16 also show that slow-wave CPW structures with multiple grounded substrate shield layers gain a slight reduction in phase velocity at the cost of a large reduction in the peak *Q* (Fig. 15).

### F. A General Slow-wave CPW Design Methodology

A general methodology can be formulated from the analyses and results above for the design of slow-wave CPW devices in CMOS with maximum peak *Q* values.

The first step is to choose a suitable length and width for the device to effectively exploit the slow-wave effect. Next, specific metal layers are chosen for the signal conductor, coplanar ground lines and substrate shield. The thick top metal layer should be used for the signal and coplanar ground lines to minimize losses due to series and skin effect resistances. The slow-wave effect is maximized when maximum capacitance and approximately constant inductance. This is achieved by forming the top layer of the substrate shield on the metal layer just below the signal and co-planar ground conductors.

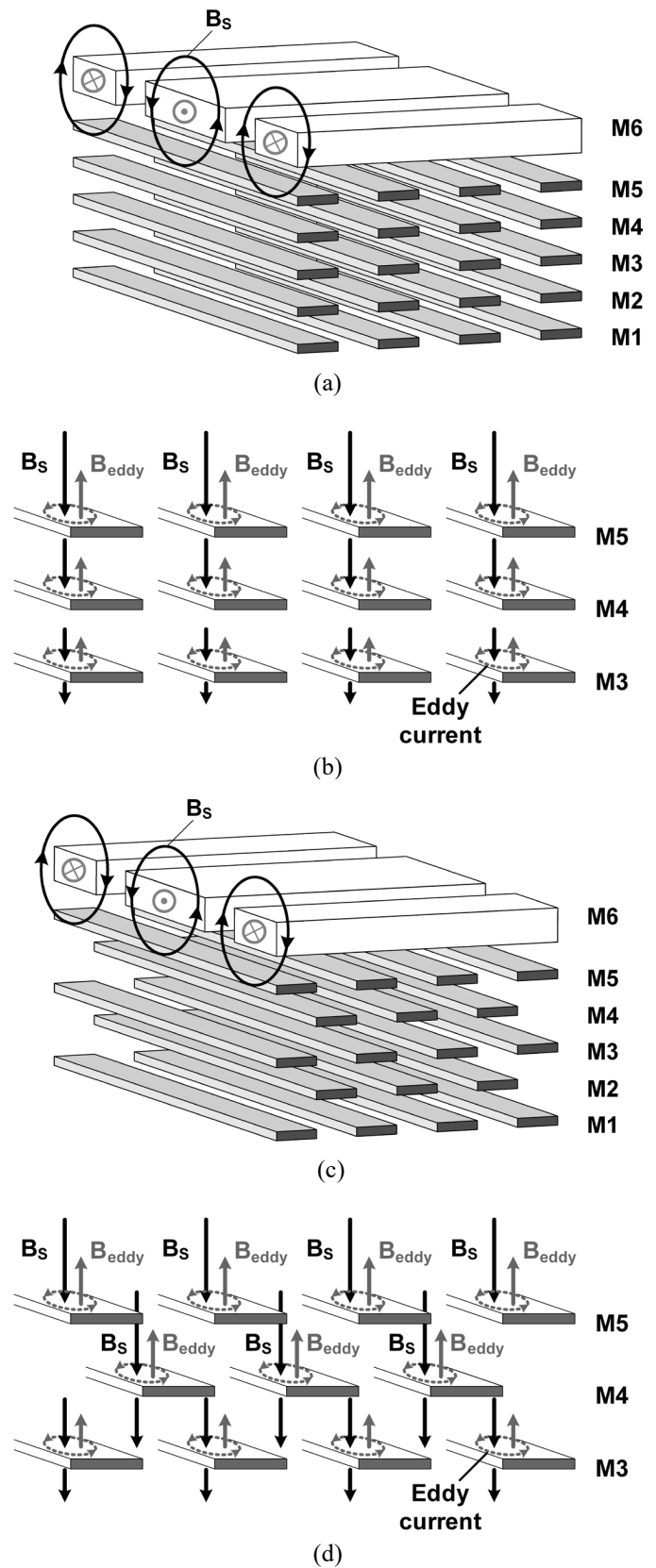

Fig. 14. Slow-wave CPW with multiple (a) overlapping and (c) offset metal substrate shield strip stacks. The corresponding magnetic flux densities and eddy currents are depicted in (b) and (d), respectively.

TABLE IV
GROUNDED AND FLOATING SHIELD Slow-wave CPW
($W$ = 180 μm, $S_L$ = 3 μm, $S_S$ = 3 μm, $Z_o \sim 40$ Ω)

| Device #<br>Device Type | Substrate Shield | $W_S$ / $S$ / $W_G$ (μm) |
|---|---|---|
| 1<br>Fig. 8(a) | Grounded | 16 / 57 / 25 |
| 2<br>Figs. 8(d) & 14(a) | Grounded | 16 / 57 / 25 |
| 3<br>Fig. 8(e) & 14(a) | Grounded | 16 / 57 / 25 |
| 4<br>Fig. 8(d) & 14(c) | Grounded | 15 / 57.5 / 25 |
| 5<br>Fig. 8(e) & 14(c) | Grounded | 15 / 57.5 / 25 |
| 6<br>Fig. 8(a) | Floating | 20 / 55 / 25 |
| 7<br>Figs. 8(d) & 14(a) | Floating | 19 / 55.5 / 25 |
| 8<br>Figs. 8(e) & 14(a) | Floating | 19 / 55.5 / 25 |
| 9<br>Figs. 8(d) & 14(c) | Floating | 18 / 56 / 25 |
| 10<br>Figs. 8(e) & 14(c) | Floating | 18 / 56 / 25 |
| 11<br>(*M5*: Grounded)<br>(*M4*-1: Floating)<br>Fig. 14(a) | Grounded / Floating | 15 / 57.5 / 25 |
| 12<br>(*M5*: Grounded)<br>(*M4*-1: Floating)<br>Fig. 14(c) | Grounded / Floating | 15 / 57.5 / 25 |

For a desired $Z_o$, $W_S$, $W_G$ and $S$ are found in conjunction with $S_L$ and $S_S$; the structure of Fig. 8(a) is a preferred starting point. For a given $W$, $S$ is found following optimizations for $W_S$ and $W_G$ for achieving high $Q$. Recall that capacitance increases at a faster rate for increases in $W_S$ than for $W_G$. Hence, in order to use die area efficiently, $W_S$ should be weighted more heavily than $W_G$ during the optimization process. The results presented earlier show that grounded (floating) substrate shield strips are preferred for devices with low-$Z_o$ (high-$Z_o$). The results also suggest that $S_L$ should be of minimum length to minimize eddy current losses; $S_S$ should also have a small value.

Finally, after all of the dimensions are determined, different combinations of signal conductor and coplanar ground metal stacks should be considered if $Q$ using the structure of Fig. 8(a) is insufficient. The devices shown in Figs. 8(b)(c) provide a high peak $Q$, low $v_p$ and nearly constant $Z_o$ when compared to the device of Fig. 8(a).

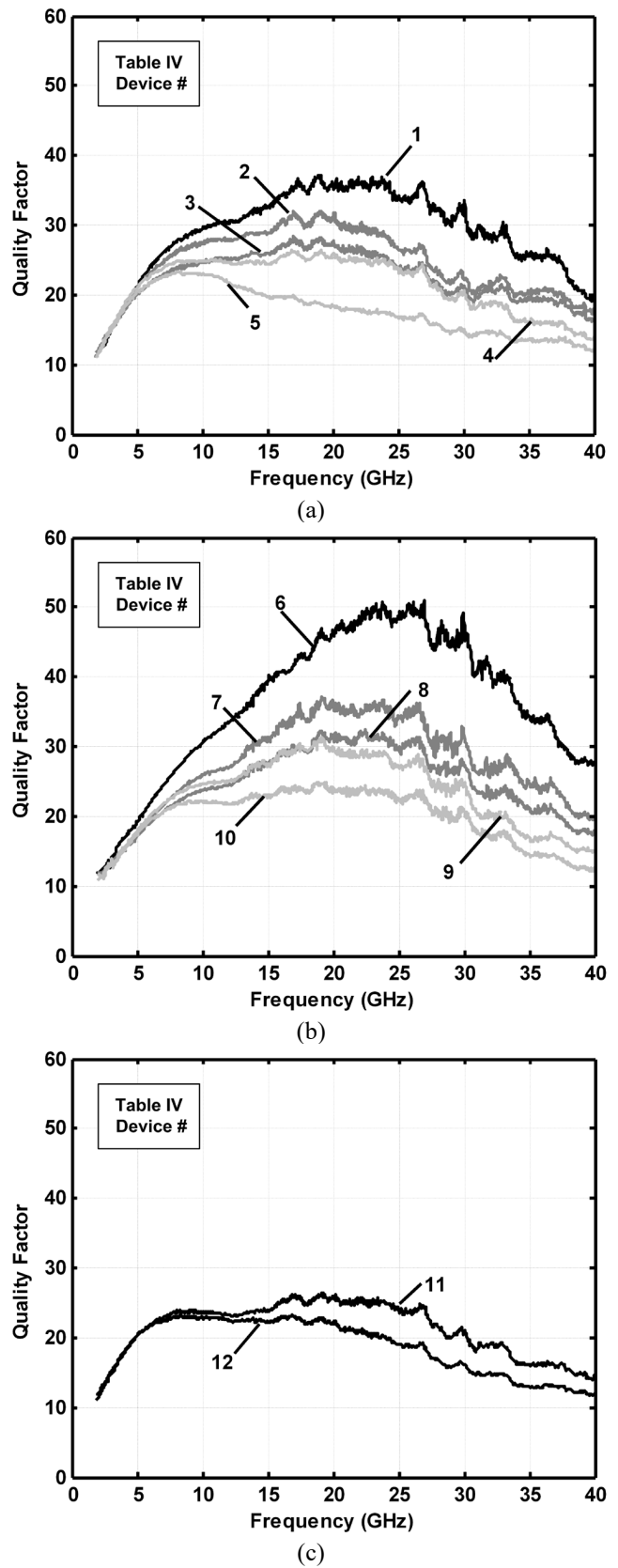

Fig. 15. The effects of grounded vs. floating vs. hybrid (i.e., grounded and floating) substrate shield strips on $Q$. Device dimensions are given in Table IV.

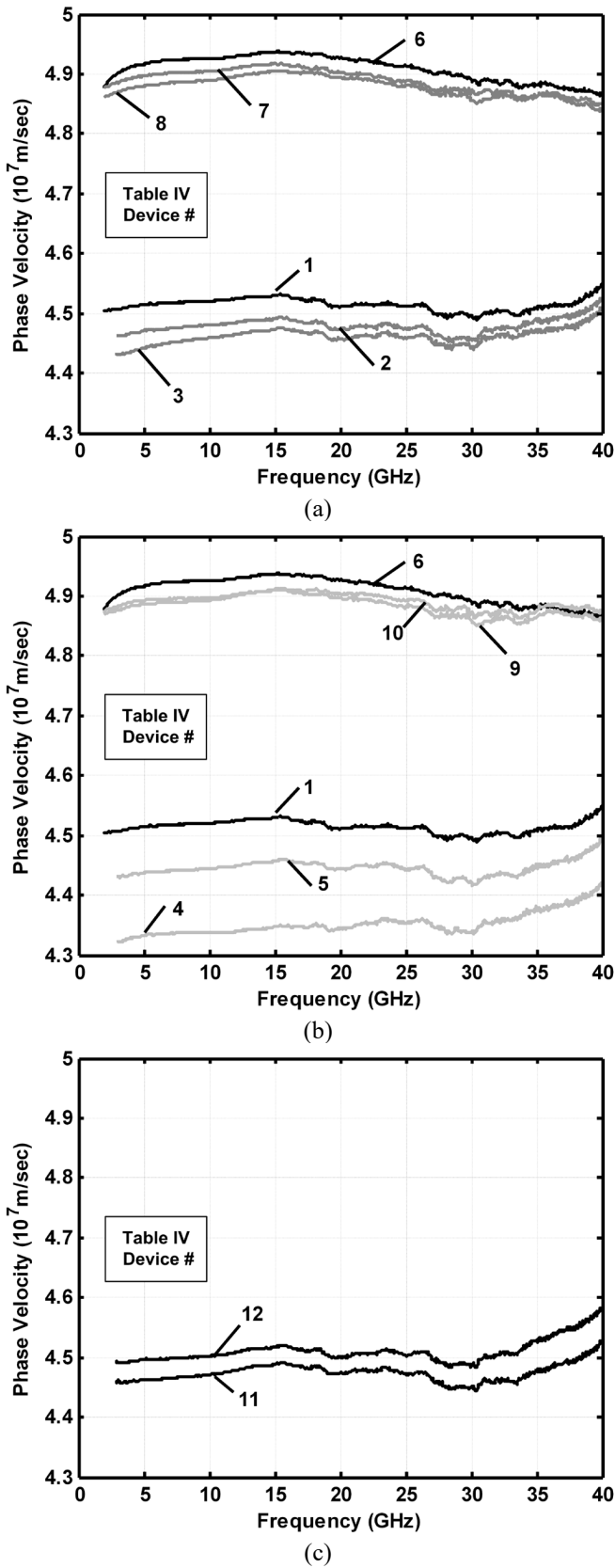

Fig. 16. The effects of grounded vs. floating vs. hybrid (i.e., grounded and floating) substrate shield strips on $v_p$. Device dimensions are given in Table IV.

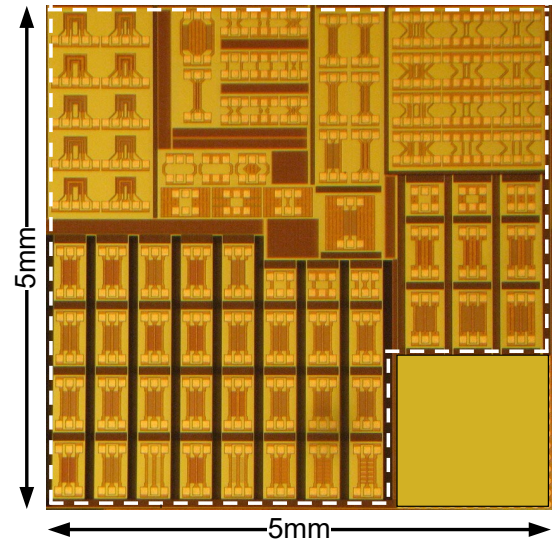

Fig. 17. The experimental chip in a 6-metal 0.18 μm CMOS process.

Fig. 17 shows a microphotograph of the 5 mm x 5 mm die in a 6-layer 0.18 μm CMOS process. It comprises about 50 different conventional, slow-wave and U-shaped slow-wave CPW [25] devices. As mentioned earlier, several on-wafer SOLT de-embedding structures are included. An Agilent HP8364A network analyzer was used to measure the scattering parameters, which were transformed to determine the parameter values for the transmission lines [26]. About 10 die measured from the same wafer with SOLT recalibration for each measurement showed consistent results.

## IV. CONCLUSIONS

A comprehensive experimental study of slow-wave coplanar waveguide devices is presented. A general methodology enables high-$Q$ and low $v_p$ designs for a given $Z_o$. $Q$- optimization is detailed in terms of the widths of the signal and coplanar ground lines, the spacing between them, the length and spacing of the substrate shield strips, the effective thickness of the signal and coplanar ground metal stacks and substrate shield metal stacks and their placements on the different metal layers.

Several conclusions are drawn from this study:

(1) The preferred slow-wave CPW comprises thick signal and coplanar ground lines with a single substrate shield layer just below.
(2) Greater overall $W$ gives greater design flexibility; however, there exists an optimum $W$ for highest $Q$.
(3) Minimum $S_L$ and $S_S$ are recommended for low-$Z_o$ designs. $S_S$ can be increased slightly for a device with a high $Z_o$.
(4) Grounded (floating) substrate shields are preferred for low-$Z_o$ (high-$Z_o$) designs.

A key conclusion from this research is that the achievable maximum peak $Q$ value is increased by 7X when the optimum topology with optimum device dimensions is used; $v_p$ is also reduced by 1.5X at 24 GHz. The reduced die area and high-Q performance make slow-wave co-planar waveguides a viable option for future RF applications in CMOS.